\documentclass[twocolumn]{aastex63}
\usepackage{amsmath}
\usepackage{graphicx}
\usepackage{color}
\usepackage{times}
\usepackage{inputenc}
\usepackage{bm}
\usepackage{multirow}
\usepackage{url}

\def\Msun{\,M_{\odot}}
\def\fm3{\;\text{fm}^{-3}}

\begin{document}
\title{Bayesian inference of strange star equation of state using the GW170817 and GW190425 data}


\author{Zhiqiang Miao}
\affiliation{Department of Astronomy, Xiamen University, Xiamen, Fujian 361005, China; \textit{liang@xmu.edu.cn}}

\author{Jin-Liang Jiang}
\affiliation{Key Laboratory of Dark Matter and Space Astronomy, Purple Mountain Observatory, Chinese Academy of Sciences, Nanjing 210023, China}
\affiliation{School of Astronomy and Space Science, University of Science and Technology of China, Hefei, Anhui 230026, China}

\author{Ang Li}
\affiliation{Department of Astronomy, Xiamen University, Xiamen, Fujian 361005, China}

\author{Lie-Wen Chen}
\affiliation{School of Physics and Astronomy, Shanghai Key Laboratory for Particle Physics and Cosmology, and Key Laboratory for Particle Astrophysics and Cosmology (MOE), Shanghai Jiao Tong University, Shanghai 200240, China}

\date{\today}

\begin{abstract} 
The observations of compact star inspirals from LIGO/Virgo provide a valuable tool to study the highly uncertain equation of state (EOS) of dense matter at the densities in which the compact stars reside.
It is not clear whether the merging stars are neutron stars or quark stars containing self-bound quark matter. 
In this work, we explore the allowed bag-model-like EOSs by assuming the merging stars are strange quark stars (SQSs) from a Bayesian analysis employing the tidal deformability observational data of the GW170817 and GW190425 binary mergers. 
We consider two extreme states of strange quark matter, either in nonsuperfluid or color-flavor locked (CFL) and find the results in these two cases essentially reconcile. In particular, our results indicate that the sound speed in the SQS matter is approximately a constant close to the conformal limit of $c/\sqrt{3}$.
The universal relations between the mass, the tidal deformability and the compactness are provided for the SQSs. The most probable values of the maximum mass are found to be $M_{\rm TOV}=2.10_{-0.12}^{+0.12}~(2.15_{-0.14}^{+0.16})\Msun$ for normal (CFL) SQSs at a $90\%$ confidence level. The corresponding radius and tidal deformability for a $1.4 \Msun$ star are $R_{\rm 1.4}= 11.50_{-0.55}^{+0.52}~({11.42}_{-0.44}^{+0.52})~\rm km$ and $\Lambda_{1.4}= {650}_{-190}^{+230}~({630}_{-150}^{+220})$, respectively.
We also investigate the possibility of GW190814's secondary component $m_2$ of mass $2.59_{-0.09}^{+0.08}\Msun$ being an SQS, and find that it could be a CFL SQS with the pairing gap $\Delta$ larger than $244~\rm MeV$ and the effective bag parameter $B_{\rm eff}^{1/4}$ in the range of $170$ to $192$ MeV, at a $90\%$ confidence level.
\end{abstract}


\section{Introduction}
It is generally believed that the degree of freedom of dense matter is hadronic around the nuclear saturation density, $n_0 \approx 0.16$~fm$^{-3}$, and the color-flavor locked (CFL) state is expected to be the ground state of three-flavor quark matter at asymptotic densities~\citep{2008RvMP...80.1455A}.
The phase states of cold QCD matter at intermediate densities ($\sim1-10~n_0$) are unfortunately unknown~\citep[see discussions in, e.g.,][]{2010PhRvD..81j5021K,2014ApJ...789..127K,2018PhRvL.121t2701G}.
One key point is still unclear, i.e., whether compact stars are gravity-bound neutron stars (NSs) or indeed self-bound quark stars (QSs)?
After decades of speculation~\citep{1971PhRvD...4.1601B,1984PhRvD..30..272W},
QSs still serve as viable alternative physical model for compact stars~\citep{1985PhLB..160..181B,1990MPLA....5.2197G,2005PrPNP..54..193W,2016PhRvD..94h3010L,2018PhRvD..97h3015Z,2021PhRvL.126p2702B,2020arXiv200900942C,2021arXiv210202357T,2021arXiv210400544S}. 

The recently observed binary star merger events~\citep{2017PhRvL.119p1101A,2018PhRvL.121p1101A,2019PhRvX...9a1001A,2020ApJ...892L...3A,2020ApJ...896L..44A} have greatly promoted the study of the equation of state (EOS) of dense stellar matter (pressure $p$ as a function of the density $n$, or energy density $e$), restricting its stiffness or the degree of freedom of dense matter in the density regime achieved inside compact stars (possibly up to $\approx 8-10\,n_0$).
After the release of the gravitational-wave observational data from LIGO/Virgo, in-depth studies have been performed~\citep[see recent reviews by][]{2019PrPNP.10903714B,2020JHEAp..28...19L,2020GReGr..52..109C,2021GReGr..53...27D}.
However, the available mass and radius constraints derived from the two merger events, GW170817~\citep{2017PhRvL.119p1101A} and GW190425~\citep{2020ApJ...892L...3A}, commonly assume NS EOSs, with or without quark deconfinement phase transition in their cores~\citep[e.g.,][]{2017ApJ...850L..34B,2018PhRvL.121p1101A,2018PhRvL.120z1103M,2018PhRvL.120q2702F,2019PhRvX...9a1001A,2019PhRvL.123n1101F,2020ApJ...904..103M,2021ApJ...913...27L}.
Note that the quark matter in NS cores is not self-bound~\citep{2020ApJ...904..103M,2021ApJ...913...27L,2021ChPhC..45e5104X} and it only appears through a phase transition from hadronic matter at high densities.
Since the results assuming NSs cannot be used to study QS properties which have a sharp surface due to the self-bound feature of their EOS and that the parameterized NS EOS cannot catch this behavior (see also discussions in \citet{2020PhRvD.101d3003Z}),
an updated parameter space for QS EOSs based on the LIGO/Virgo observations is desirable. 

As an endeavor in this direction, in the present study, we perform a Bayesian analysis of the GW170817 and GW190425 data by assuming the merging stars are strange quark stars (SQSs).
We restrict ourselves to only two extreme cases of normal nonsuperfluid strange quark matter (SQM) and CFL quark matter since it is not known how far the CFL phase extends towards lower density.

This paper is organized as follows.
In Section 2, we introduce the EOS model for describing the SQSs. 
Section 3 presents the observational constraints 
employed and the Bayesian analysis method that we apply.
Our results are presented in Section 4 and summarized in Section 5.

\section{Normal and CFL SQS Models}
SQSs exist when one applies the Bodmer-Witten hypothesis, namely self-bound SQM is the physical nature of all compact stars~\citep{1971PhRvD...4.1601B,1984PhRvD..30..272W}.\footnote{Note that there are also discussions in the literature suggesting that non-SQM can have a lower bulk energy per baryon than normal nuclei and SQM~\citep[e.g.,][]{2018PhRvL.120v2001H,2020arXiv200900942C,2021PhRvD.103f3018Z}.}
Various approaches have been attempted for a model description of self-bound quark matter~\citep{2021ChPhC..45e5104X} since it is not attainable directly by solving QCD~\citep{2010PhRvD..81j5021K,2014ApJ...789..127K,2018PhRvL.121t2701G}.
Among them, the MIT bag model is the most widely used~\citep[see some of the latest studies, e.g.,][]{2020arXiv200912571L,2021PhRvD.103h3015R,2021PhyS...96f5302L,2021JHEAp..30...16J}. Others include the Nambu-Jona-Lasinio (NJL) model~\citep{2005PhR...407..205B}, the density-dependent quark masses~\citep{2010MNRAS.402.2715L,2011RAA....11..482L}, a confining quark matter model~\citep{1998PhLB..438..123D,2017PhRvD..96h3019C}, an interacting quark-matter model~\citep{2021PhRvD.103f3018Z} and the vector interaction enhanced-bag model~\citep{2015ApJ...810..134K}, as an incomplete list.

The SQM is composed of up ($u$), down ($d$) and strange ($s$) quarks with the charge neutrality maintained by the inclusion of electrons.
In our calculations below, we shall adopt the bag model with and without superfluidity.
For ease of discussion later, we first briefly recall some of the basic formulae and definitions.
The expressions for the grand canonical potential per unit volume in the MIT bag model is written as~\citep{1986ApJ...310..261A,1986A&A...160..121H,2005ApJ...629..969A}:
\begin{equation}
\Omega=\sum_{{i=u,\ d,\ s,\ e^-}}\Omega_i^0+\frac{3(1-a_4)}{4\pi^2}\mu^4+B_\mathrm{eff}+\frac{3m_s^4-48\Delta^2\mu^2}{16\pi^2}\ , \label{eq:canonicalpotential1}
\end{equation} 
where $\Omega_i^0$ is the grand canonical potential for particle type $i$ described as ideal Fermi gas. The baryon chemical potential is $\mu_{\rm B}=\mu_u+\mu_d+\mu_s$ and $\mu=\mu_{\rm B}/3$ is the average quark chemical potential. 
The total baryon number density can be expressed as $n = (n_u + n_d + n_s)/3$ with $n_i = -(\partial \Omega / \partial \mu_i)_{\rm V}$.
The effective bag constant $B_{\rm eff}$, accounting for the QCD vacuum's contributions, is usually regarded as a phenomenological parameter. 
The parameter $a_4$ characterizes the QCD corrections due to gluon-mediated interactions between quarks~\citep{2001PhRvD..63l1702F,2005ApJ...629..969A,2016MNRAS.457.3101B,2017ApJ...844...41L}, with $a_4 = 1$ corresponding to no QCD corrections (Fermi gas approximation).
$\Delta$ is the CFL pairing gap of an order of tens to 100 MeV~\citep{1998PhLB..422..247A,1998PhRvL..81...53R}. 
$\Delta=0$ corresponds to the nonsuperfluid case. 
There may be other superfluid phases in the intermediate densities~\citep{2017PhRvL.119p1104A,2021PhRvD.103f3018Z}, e.g., the two-flavor color-superconducting phase, a gapless CFL phase, which are not considered here and will be studied in future.
Finite quark mass was shown to have a trivial influence on the results~\citep{2020arXiv200912571L,2018PhRvD..97h3015Z}, and we neglect the masses of the up and down quarks while fix the strange quark mass as $m_s =100\,{\rm MeV}$.
The EOS of SQM, which is characterized by the independent parameters $B_{\rm eff}$ and $a_4$ ($\Delta$ as well if quark superfluity is included), can be calculated using basic thermodynamic relations, and the stars' global properties (mass $M$, radius $R$, tidal deformability $\Lambda$) can then be obtained~\citep[see details in e.g.,][]{2018PhRvD..97h3015Z,2018ApJ...862...98Z,2020ApJ...904..103M}.

\section{Observational Constraints and Bayesian Analysis}

\subsection{Bayesian Analysis}
Considering the maximal spin observed in Galactic pulsars, \citet{2019PhRvX...9a1001A} inferred that the 90\% credible intervals for the component masses of the GW170817 event lie between $1.16$ and $1.60\Msun$, and \citet{2020ApJ...892L...3A} reported that the corresponding component masses range from $1.46$ to $1.87 \Msun$ for the GW190425 event. 
We treat both events as SQS-SQS mergers. By exploiting their tidal deformability measurements, we use Bayesian inference to determine posteriors on the EOS parameter spaces of normal and CFL SQSs. 

Assuming that the noise in the LIGO/Virgo detectors is Gaussian and stationary, the likelihood of a gravitational-wave event used to perform Bayesian inference is often expressed as
\begin{equation}
\label{eq:Likelihood}
L(d|\vec{\theta}_{\rm GW})\!\propto\!{\rm Exp}\!\left(-2 \int \frac{|d(f) - h(\vec{\theta}_{\rm GW},f)|^2}{S_{\rm n}(f)} df\right),
\end{equation}
where $S_{\rm n}(f)$, $d(f)$, and $h(\vec{\theta}_{\rm GW},f)$, respectively, denote the power spectral density (PSD), the frequency domain data, and the frequency domain waveform generated using parameter set $\vec{\theta}_{\rm GW}$. The tidal deformability $\Lambda$ encoded in the gravitational-wave strain data can be mapped from the mass through the EOS. Thus the EOS parameters together with component masses can be incorporated to construct the gravitational-wave parameters $\vec{\theta}_{\rm GW}$~\citep[e.g.,][]{2020ApJ...892...55J,2020ApJ...888...45T}. 
We take the publicly available strain data\footnote{\url{https://www.gw-openscience.org/eventapi}} and PSDs\footnote{The PSD of GW170817 can be found at \url{https://doi.org/10.7935/KSX7-QQ51}, while for GW190425, the PSD can be found in the parameter estimation sample release at \url{https://dcc.ligo.org/LIGO-P2000026/public}.} of GW170817 and GW190425, together with the waveform model {\sc IMRPhenomD\_NRTidal} \citep{2017PhRvD..96l1501D} to do the analysis. 
We follow all the data analysis details of GW170817 and GW190425 described in \citep{2019PhRvX...9a1001A} and \citep{2020ApJ...892L...3A}, respectively, except that we only consider the low-spin case and do not consider the calibration error which causes a minor difference in our results. 
When combining the two events with the analysis, we encounter a problem whereby many sampling parameters make it computationally expensive to convergence in the Nest sampling algorithm. So we take the interpolated likelihood tables of \citet{2020MNRAS.499.5972H} into the analysis, which marginalized over all the other parameters except masses and tidal deformabilities. We find that using the interpolated likelihood tables is well consistent with those of using gravitational-wave data in analyzing a single event. Then by employing the python-based {\sc Bilby} \citep{2019ApJS..241...27A} and {\sc Pymultinest} \citep{2016ascl.soft06005B} packages, we simultaneously infer the gravitational-wave parameters and the EOS parameters.

\subsection{Priors and Constraints}

To improve the converging rate of the nest sampling, we marginalize the coalescence phase parameter in the likelihood and fix the source's sky location determined by the electromagnetic observations \citep{2017ApJ...848L..12A,2017ApJ...848L..28L}. As for the priors of the other parameters in $\vec{\theta}_{\rm GW}$, we take a similar choice as presented in~\citet{2020ApJ...888...45T}. For the parameters $\vec{\theta}_{\rm EOS}$ that construct the EOS of SQSs, following our previous studies~\citep{2018PhRvD..97h3015Z}, we assign wide boundaries to them as $B_{\rm eff}^{1/4}\in[125, 150]\,{\rm MeV}$, $\Delta \in[0, 100]\,{\rm MeV}$, and $a_4\in[0.4, 1]$ as theoretically estimated, with which both uniform and logarithmic uniform distributions are investigated. For technical reasons, the lower bound of the logarithmic uniform distribution cannot be zero; thus, we set a reasonable lower bound $0.1$ for $\Delta$ in the logarithmic uniform case.

As mentioned earlier, we assume all compact stars are self-bound SQSs and that they are composed of charge-neutral bulk SQM.
Consequently, two stability constraints should be adopted: first, the energy per baryon for non-SQM should satisfy $(E/A)_{\rm ud}\geq934\,{\rm MeV}$ to guarantee the observed stability of atomic nuclei; second, $(E/A)_{\rm uds}\leq930\,{\rm MeV}$ is required, according to the hypothesis that SQM is absolutely stable~\citep{1971PhRvD...4.1601B,1984PhRvD..30..272W}. see also~\citet{2010MNRAS.402.2715L,2011RAA....11..482L}.
In addition, the causality condition for the SQM EOS is guaranteed for all the SQS calculations. 
The mass measurement of massive pulsars establishes a lower bound on the maximum mass of SQSs. 
Only the EOSs that support a $M_{\rm TOV}$ larger than this lower bound can pass this constraint, while others will be rejected.
We adopt the largest mass measured through Shapiro delay, $M=2.08\pm0.07\Msun$ (68\% confidence level) of MSP J0740+6620~\citep{2020NatAs...4...72C,2021ApJ...915L..12F}, to place the $M_{\rm TOV}$ constraint. 
A Gaussian-like likelihood is used to encapsulate the mass measurement of the $2.08\,\Msun$ pulsar. More details on the implementation of the $M_{\rm TOV}$ limit can be found in our previous study~\citep{2021ApJ...913...27L}.

\section{Results and Discussion}

\begin{figure*}
        \includegraphics[width=3.3in]{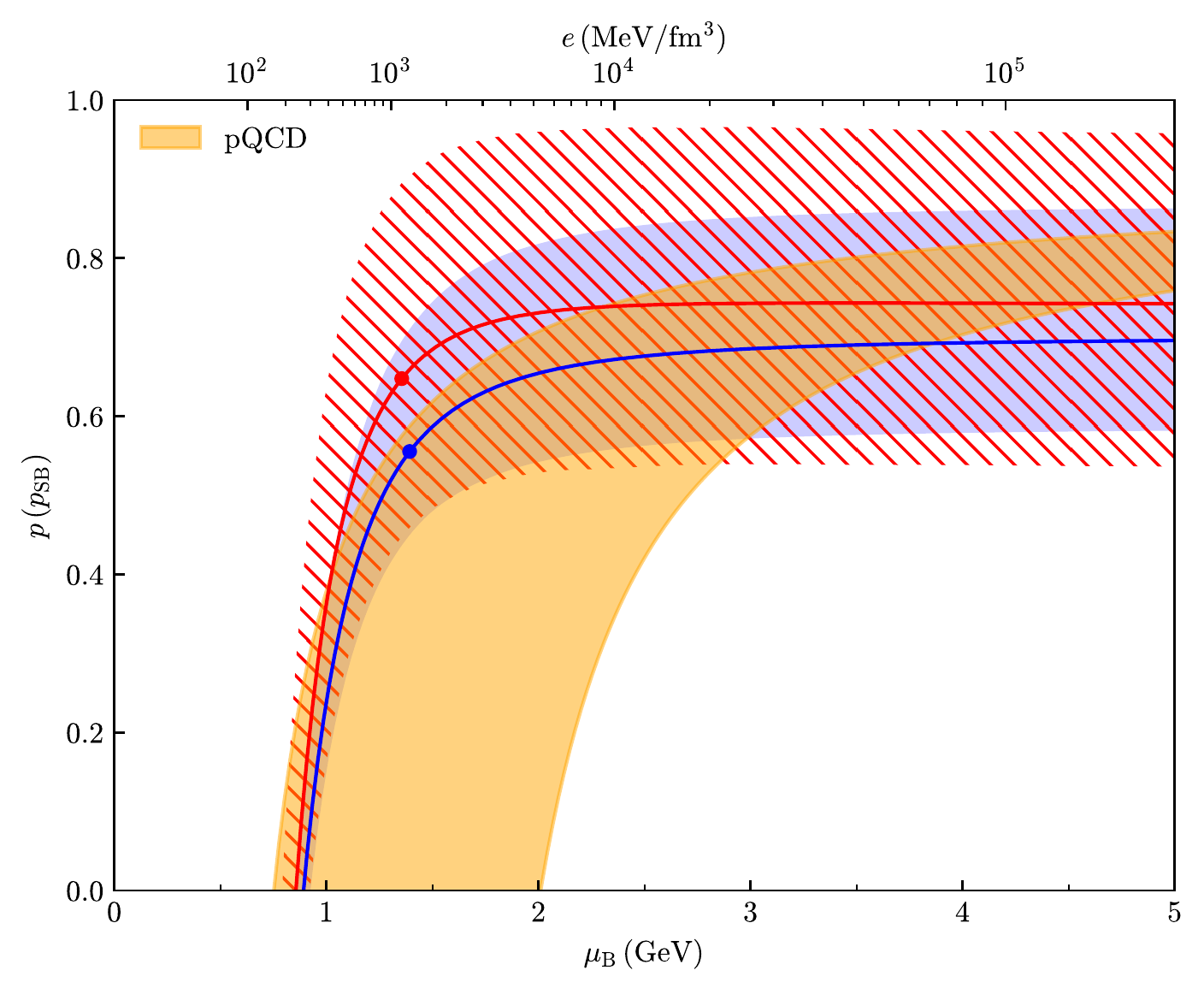}
        \includegraphics[width=3.3in]{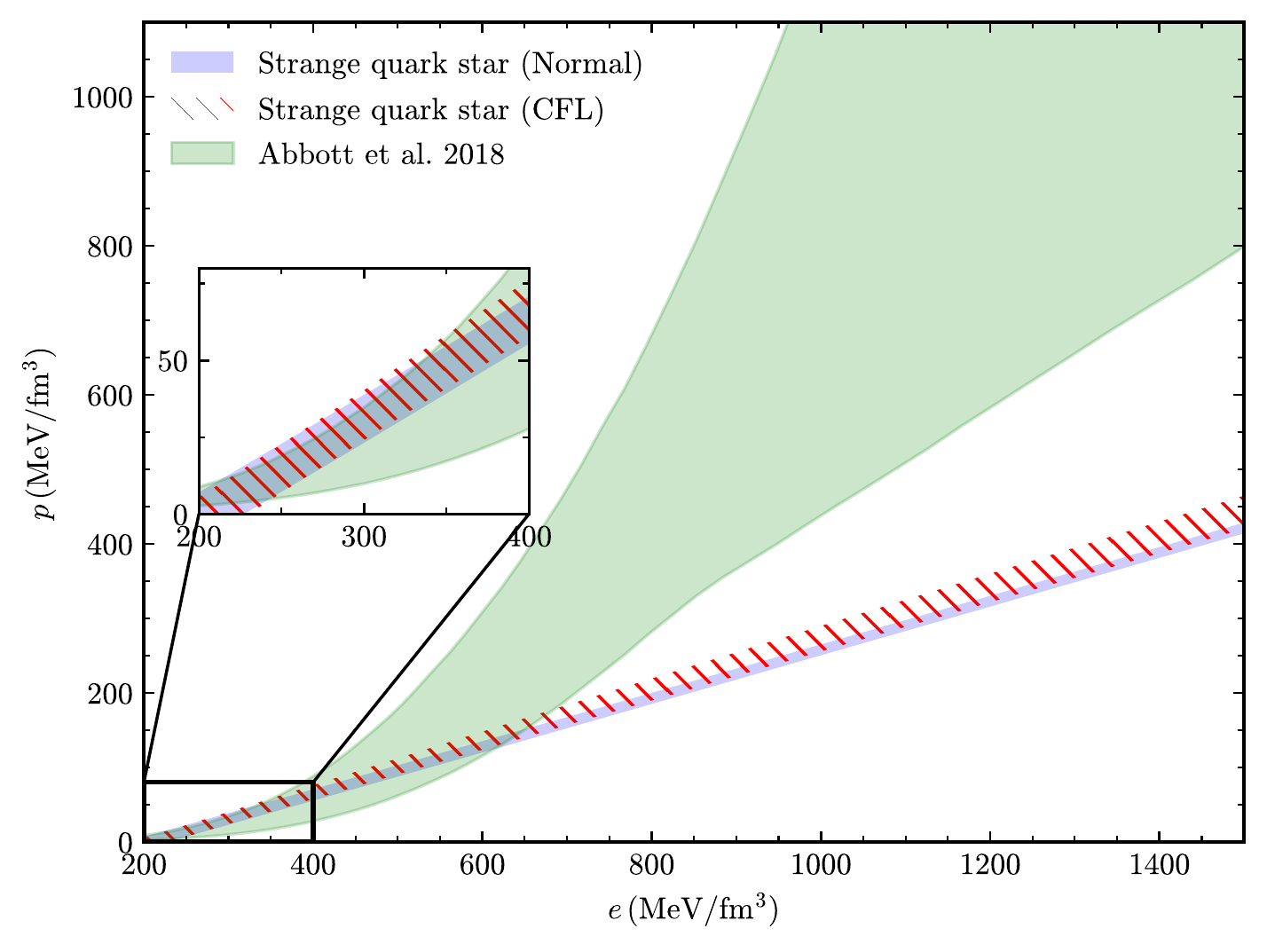}
    \caption{Posterior distributions ($90\%$ confidence level) of the pressure vs. chemical potential relation (left panel) and the EOS (right panel) for normal and CFL QSs from the GW170817 and GW190425 data, together with the pQCD result from \citet{2010PhRvD..81j5021K} in the left panel and the NS result based on GW170817 from LIGO/Virgo~\citep{{2018PhRvL.121p1101A}} in the right panel.
     The central densities of normal and CFL maximum-mass stars are marked in the left panel with blue and red symbols.
    An additional upper horizontal axis are indicated in the left panel, showing the corresponding energy density within the most favored EOS parameter sets in the QS analysis.
    }
    \label{fig:eos}
\end{figure*}
\subsection{EOS and the Pressure vs. Chemical Potential Relation}

\begin{table}
  \centering
  \caption{Most probable intervals of the EOS parameters ($90\%$ confidence level) constrained by the joint GW170817+GW190425 analysis for the two priors: Uniform ($U$) and logarithmic uniform (${\rm log}U$) distributions.
  }
    \setlength{\tabcolsep}{0.8pt}
\renewcommand\arraystretch{1.3}
\begin{ruledtabular}
\begin{tabular*}{\hsize}{@{}@{\extracolsep{\fill}}lccc@{}}
   Parameters & Prior Type & 
   & Joint Analysis \\ 
    \hline 
    $B_{\rm eff}^{1/4}/\rm MeV$ 
              & $U(125, 150$) & Normal  
              &  $137.6_{-4.4}^{+4.8}$ \\ 
              &               & CFL   
              &  $144.3_{-7.9}^{+5.0}$ \\
    & ${\rm log}U(125, 150$)& Normal 
    &  $137.3_{-4.2}^{+4.7}$ \\
    &                 & CFL    
    &  $138.5_{-5.3}^{+8.3}$ \\ 
    \hline 
     $a_4$    
    & $U(0.4, 1$)   & Normal 
    &  $0.70_{-0.12}^{+0.16}$ \\ 
    &               & CFL
    &  $0.74_{-0.20}^{+0.22}$ \\ 
    & ${\rm log}U(0.4, 1$) & Normal
    &  $0.69_{-0.11}^{+0.15}$ \\ 
    &                & CFL
    &  $0.70_{-0.13}^{+0.18}$ \\ 
    \hline 
     $\Delta/\rm MeV$ 
      & $U(0, 100$)      & CFL 
      &  ${67.6}_{-54.3}^{+28.1}$ \\ 
      & ${\rm log}U(0.1, 100$) & CFL 
      &  $6.6_{-6.4}^{+73.7}$ \\
  \end{tabular*}
  \end{ruledtabular}
  \label{tb:priors_posts}
\end{table}

\begin{figure*}
         \includegraphics[width=2.3in]{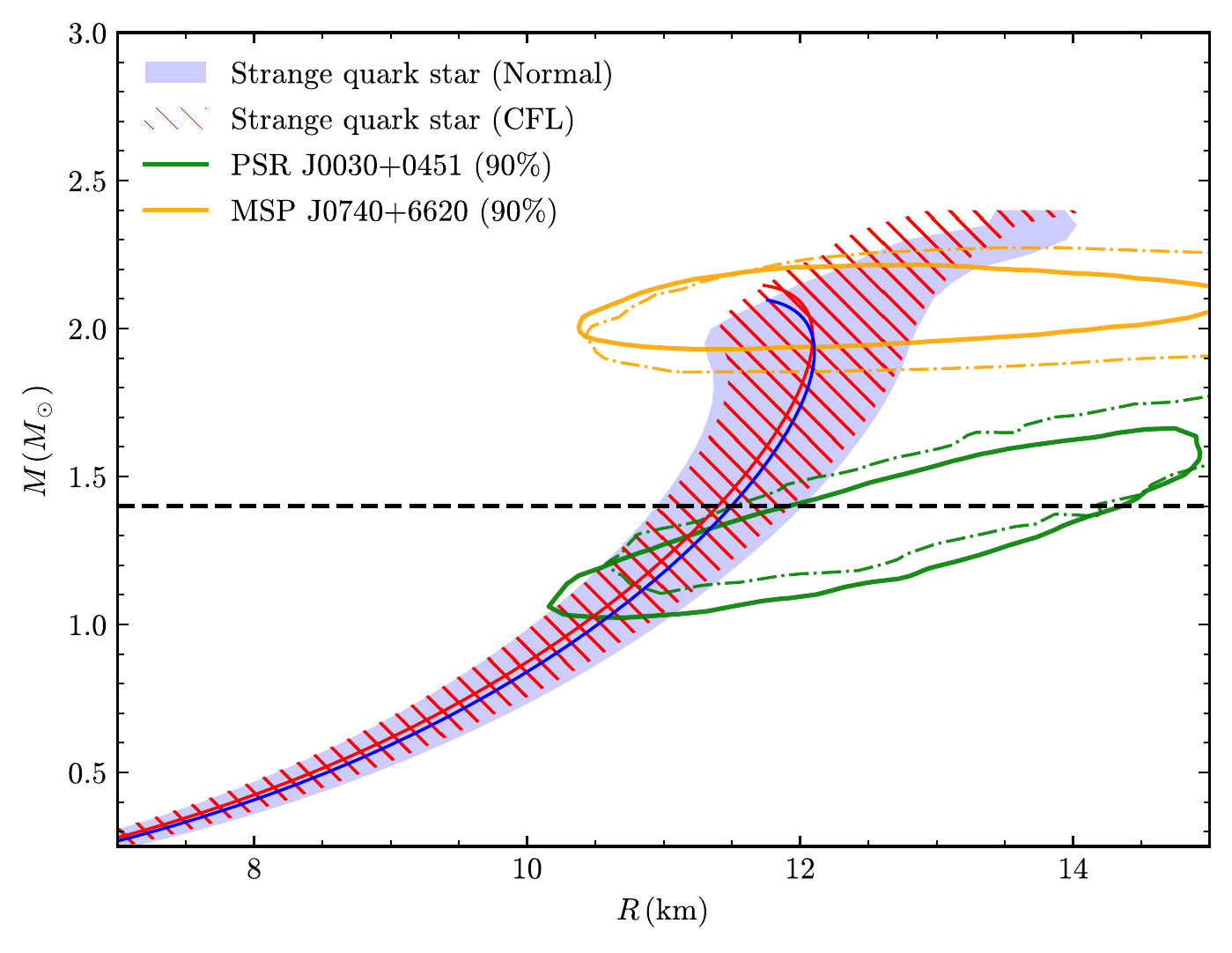}
            \includegraphics[width=2.3in]{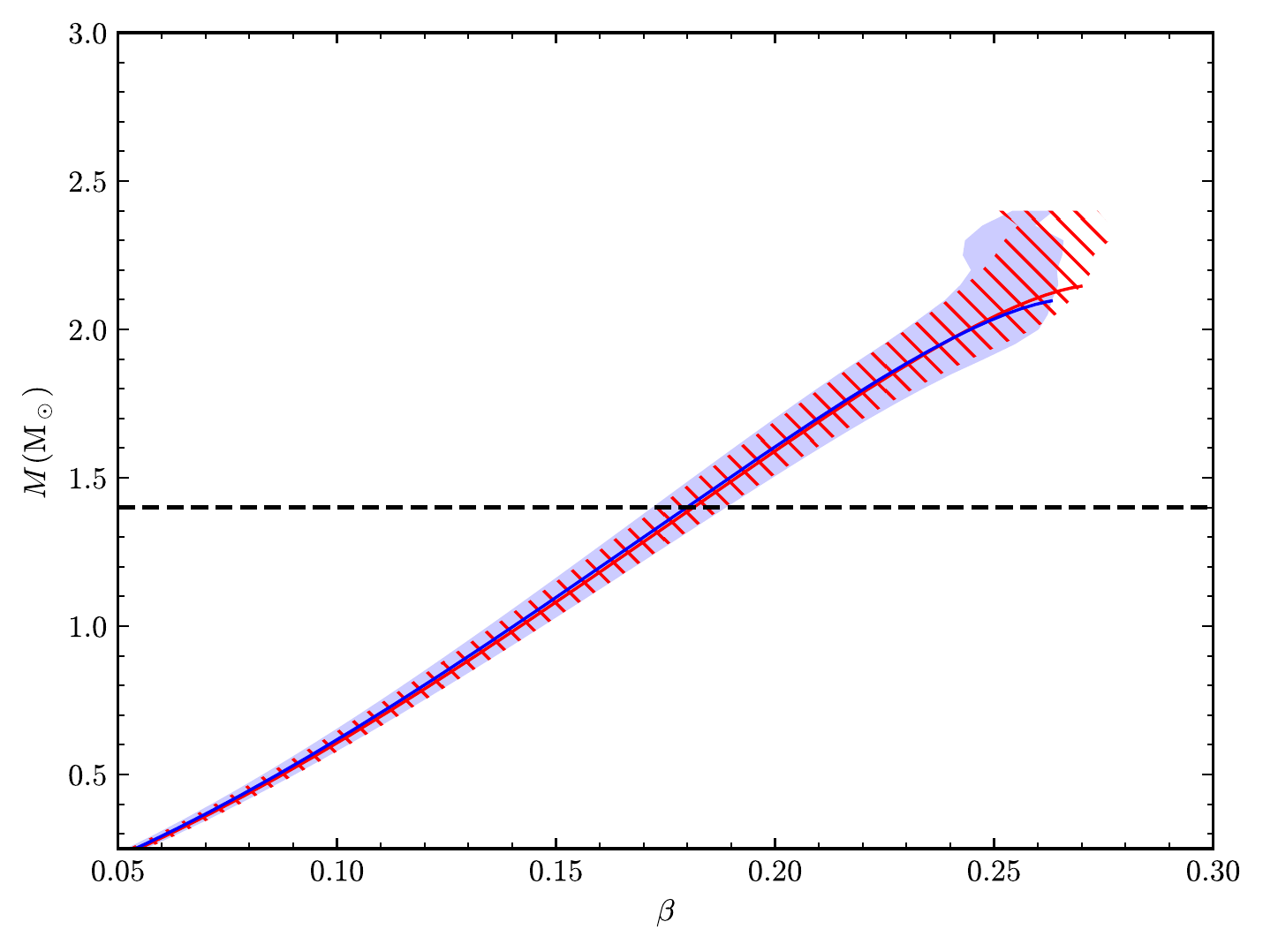}
        \includegraphics[width=2.3in]{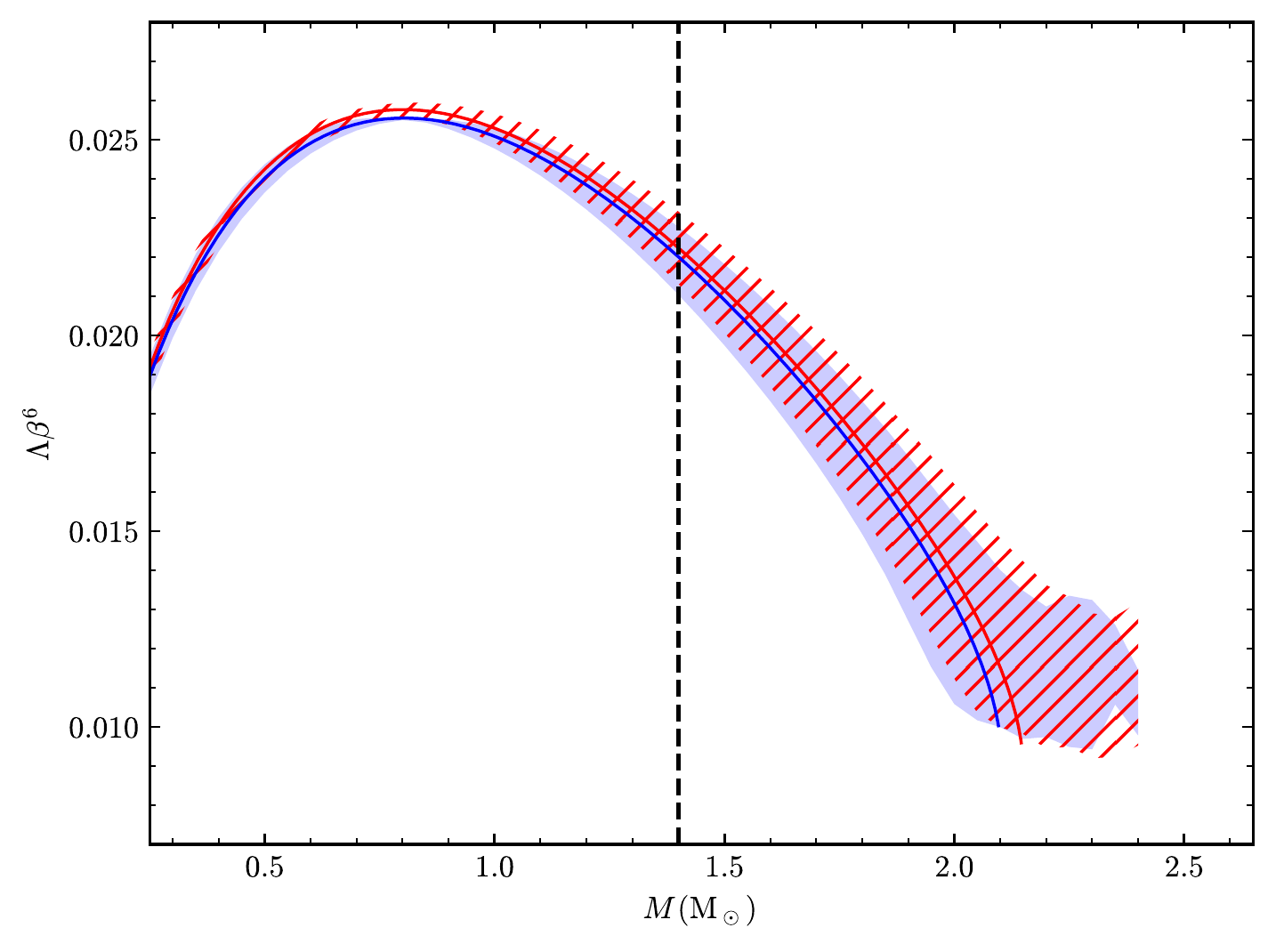}
    \caption{Posterior distributions of the mass-radius relation (left), the mass-compactness relation (middle) and $\Lambda \beta^6$ (right) as functions of the mass for normal and CFL SQSs from the joint GW170817 and GW190425 analysis, to the 90\% confidence level.
     Also shown in the left panel are the mass and radius analysis of PSR J0030+0451 and MSP J0740+6620 based on NS EOSs from the Neutron Star Interior Composition Explorer~\citep{2019ApJ...887L..24M,2019ApJ...887L..21R,2021arXiv210506980R,2021arXiv210506979M}. 
    }
    \label{fig:lambda}
\end{figure*}

\begin{figure}
    \includegraphics[width=3.3in]{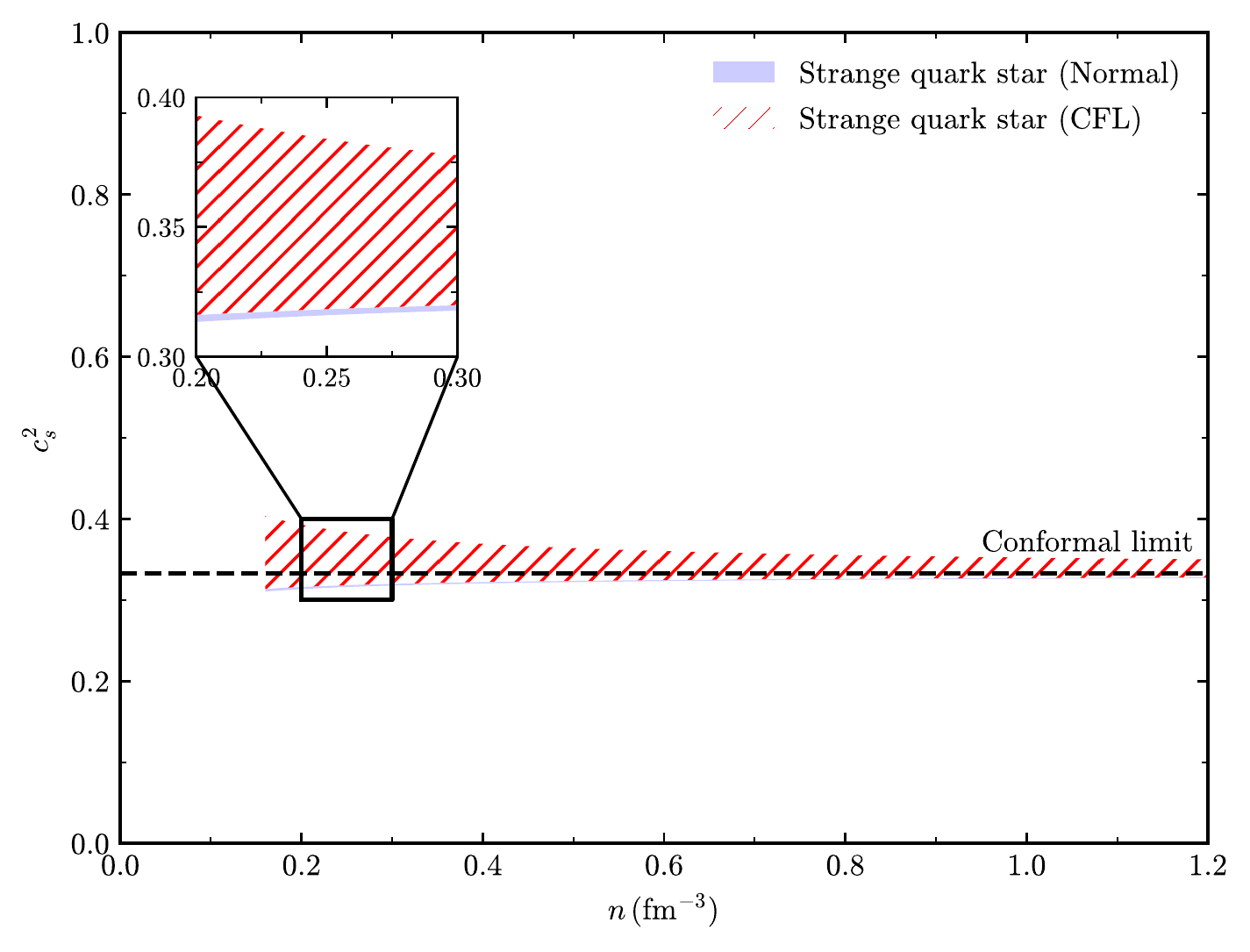}
    \caption{Posterior distributions of the squared sound speed $c^2_{\rm s}$ (in units of $c^2$) as a function of the density, for normal and CFL SQSs from the joint GW170817 and GW190425 analysis, to the 90\% confidence level. Also shown is the conformal limit of $c/3$.} 
    \label{fig:cs}
\end{figure}

The most probable values of the EOS parameters and their $90\%$ confidence boundaries, constrained jointly by the GW170817 and GW190425 data, are reported in Table \ref{tb:priors_posts}. 
The corresponding results of the EOS (including the sound velocity) and the SQS properties are shown in Figs.~\ref{fig:eos}-\ref{fig:cs}.

From Table \ref{tb:priors_posts}, we see that two EOS parameters ($B_{\rm eff}^{1/4}$ and $a_4$) are relatively well constrained: 
$B_{\rm eff}^{1/4}\approx137~\rm MeV$ ($144~\rm MeV$) and $a_4\approx0.70$ ($0.74$) for the normal (CFL) matter, which depend weakly on the prior choice.
It is worth mentioning that the ``standard value'' of the bag parameter is $(144\,\rm MeV)^4$ ($56\,\rm MeV/fm^3$) from reproducing the mass spectrum of light hadrons and heavy mesons~\citep{1975PhRvD..12.2060D,1980PhRvD..22.1198H}, and the estimated value from lattice calculations at zero chemical potential is $\sim(212\,\rm MeV)^4$ ($\approx262\,\rm MeV/fm^3$)~\citep{1996NuPhA.606..320B}.
The inferred $a_4$ values here are close to the value suggested in~\citet{2001PhRvD..63l1702F}.
However, the color superconductivity gap $\Delta$ is poorly constrained by the observed global star properties.
Nevertheless, the $\Delta$ could be constrained through future tidal deformability measurement of massive QSs (close to the maximum mass) as discussed in~\citet{2020arXiv200912571L}.

In Fig.~\ref{fig:eos}, we report the posterior distributions of the SQS EOS as well as the pressure vs. chemical potential ($p$-$\mu_B$) relation. 
Also shown in the $p$-$\mu_{\rm B}$ plot is the results from perturbative QCD~\citep{2010PhRvD..81j5021K}, which is only applicable above $\mu_{\rm B}\sim2.6~\rm GeV$.
It is seen that the observational data can constrain the EOS effectively in the nonperturbative realm where the properties of compact stars are relevant. 
It is common in the literature to analyze the gravitational-wave data by assuming all compact stars are NSs; the resulting EOS from LIGO/Virgo~\citep{2018PhRvL.121p1101A} is included in the right panel of Fig.~\ref{fig:eos}.
It is seen that SQS EOSs become stiff earlier than the NS ones~\citep{2018PhRvL.121p1101A} at low densities, indicating the sharp surface of SQSs is of a relatively low density~\citep{2017ApJ...844...41L}, as low as $\sim0.22~\rm fm^{-3}$ ($\approx1.4n_0$), to the $90\%$ confidence level.
However, at high densities, SQS EOSs are softer than the NS ones~\citep{2018PhRvL.121p1101A,2021ApJ...913...27L}.

\subsection{SQS Mass, Radius and Tidal deformability}

Generally, in the bag model, the stability of quark matter is dominated by the vacuum term ($B_{\rm eff}$) and the perturbative interaction term ($a_4$)~\citep{1984PhRvD..30.2379F}, with the quark pairing $\Delta$ helping to lower the energy~\citep{2001PhRvD..64g4017A,2001PhRvL..86.3492R,2002PhRvD..66g4017L,2003A&A...403..173L,2018PhRvD..97h3015Z}.
As a consequence, the introduction of pairing (as an extra degree of freedom) allows wider parameter spaces for the model parameters ($B_{\rm eff}, a_4$), the EOS, and SQS properties, as one can observe in Table \ref{tb:priors_posts} and Figs.~\ref{fig:eos}-\ref{fig:cs}.
In particular, the CFL matter (described in a three-parameter model) allows more parameter space for SQS EOSs than the normal SQM matter (described in a two-parameter model)~\citep{2020arXiv200912571L}. 
However, their predictions for SQS mass, radius, and tidal deformability agree considerably well with each other, as demonstrated in Fig.~\ref{fig:lambda}. 
As a result, the SQS's mass and tidal deformability can be described nearly universally as functions of the mass/radius (or equivalently the compactness $\beta \equiv M/R$): $M(\beta)=13.07\times\beta^{1.32}$ and $\Lambda\beta^6(M)=0.02\times M^{2.25}/(\exp(0.53\times M^{1.88})-1)$, with their coefficient of determination being $r^2=0.993$ and $r^2=0.989$; 
One can also evaluate the stars' important properties, for example, the maximum mass is $M_{\rm TOV}=2.10_{-0.12}^{+0.12}~(2.15_{-0.14}^{+0.16})\Msun$, the radius for a $1.4 \,M_{\odot}$ star is $R_{\rm 1.4}= 11.50_{-0.55}^{+0.52}~({11.42}_{-0.44}^{+0.52})~\rm km$ and the corresponding tidal deformability is $\Lambda_{1.4}= {650}_{-190}^{+230}~({630}_{-150}^{+220})$ for normal (CFL) SQSs, at a $90\%$ confidence level.
The maximum mass around $2.15\Msun$ (with an upper bound of $2.31\Msun$) is very close to our previous theoretical calculations~\citep{2018PhRvD..97h3015Z} with only the GW170817 constraint considered.
These results are potentially useful for the discussion of the possible transition from NSs to QSs in the two-family scenario~\citep{2007ApJ...659.1519D,2009PhLB..680..448B,2014PhRvD..89d3014D,2019ApJ...881..122D,2019AIPC.2127b0026C,2020PhRvD.102f3003D}, where NSs coexist with QSs and may transit to QSs in their evolution as the central engines of short gamma-ray bursts~\citep{1996PhRvL..77.1210C,2000ApJ...530L..69B,2016PhRvD..94h3010L,2017ApJ...844...41L}.

\subsection{Sound Speed in Dense Matter}

The earlier EOS stiffening in SQSs, compared to in NSs, has profound influences on the density behavior of the squared speed of sound $c^2_s$ (in units of $c^2$) in the two kinds of stellar matter, which is shown in Fig.~\ref{fig:cs}.
There have been many previous discussions saying that, to fulfill the two-solar-mass lower-mass limit for compact stars, the $c_s$ value in NS matter should substantially exceed the conformal limit of $c/\sqrt{3}$~\citep{2014ApJ...789..127K,2015PhRvL.114c1103B,2018MNRAS.478.1377A}, even close to $\sim0.9c$ in some studies~\citep{2018PhRvC..98d5804T}, at around $5n_0$. 
Such requirements are confirmed by both theoretical calculations~\citep{2021ChPhC..45e5104X} and statistical analysis~\citep{2021ApJ...913...27L,2020PhRvD.101l3007L}.
Considerable efforts are undertaken in the literature to pursue the underlying mechanism explaining the rapid growth in $c_s$~\citep{2019PhRvL.122l2701M}.
However, from our present study of SQSs with a bag-model-like EOS, it is found that $c_s$ is essentially a constant close to $c/\sqrt{3}$~\citep[see also ][]{2020ApJ...905....9T,2021arXiv210202357T}); 
A heavy compact star does not necessarily demand a superconformal $c_s$ as long as the EOS stiffening happens early.
In fact, we generally have two kinds of EOSs in our model:
the first kind has a maximum sound speed at a low density corresponding to the zero-pressure point, while the second kind reaches its maximum $c/\sqrt{3}$ at asymptotic density.
This is related to the uncertain superfluid phases mentioned above, controlled in the grand canonical potential by the sign of a combined term, $m_s^2-4\Delta^2$~(named coefficient $a_2$ in~\citet{2005ApJ...629..969A}). 
A negative $a_2$ corresponds to the first kind while a positive one corresponds to the second kind~\citep[see more discussions in][]{2005ApJ...629..969A,2021PhRvD.103f3018Z}.
This causes the different $c_s$ behavior in the normal and CFL cases, although they both give an approximately constant sound speed close to the conformal limit.
We mention here that the QS study from gravitational wave data is still at an early stage with simple EOS modeling, and it will be interesting to see how our results change if more sophisticated quark-matter models are applied. For instance, it will be fascinating to explore SQSs in the future by adopting the vector interaction enhanced-bag model~\citep{2015ApJ...810..134K}, where the effects of dynamical chiral symmetry breaking and vector repulsion are included, based on the NJL model with vector interaction.

\subsection{Studies on CFL SQSs with Enlarged Parameter Spaces for Effective Bag and Superconducting Gap}

In the analysis above, the parameters of the effective bag $B_{\rm eff}$ and CFL pairing gap $\Delta$ are varied in their theoretically estimated regions, following our previous study~\citep{2018PhRvD..97h3015Z}.
However, some larger $\Delta$ values are also used in the literature~\citep{2003PhRvD..67g4024A}. 
In particular, much higher values for $B_{\rm eff}$ and $\Delta$ are used in recent strange matter studies when discussing the possibility of GW190814's secondary component $m_2$ of mass $2.50-2.67\Msun$~\citep{2020ApJ...896L..44A} as a static CFL SQS~\citep{2021PhRvD.103h3015R}. 
It was found that this is possible with the combination of a large bag parameter ($B_{\rm eff}^{1/4}>159~\rm MeV$) and pairing parameter ($\Delta>200~\rm MeV$).

By employing the mass measurement of GW190814's secondary component, $2.59_{-0.09}^{+0.08}\Msun$ (at the $90\%$ credible level), as the lower bound on the maximum mass (instead of the MSP J0740+6620 measurement used above), we extend the ranges of both parameters and repeat the analysis. In particular, the upper boundary of $B_{\rm eff}^{1/4}$ ($\Delta$) is increased from $150~\rm MeV$ ($100~\rm MeV$) to $250~\rm MeV$ ($500~\rm MeV$).
Our results suggest $\Delta>244\,{\rm MeV}$ and $170\,{\rm MeV}<B_{\rm eff}^{1/4}<192\,{\rm MeV}$, to the 90\% confidence level, in the context of GW190814's secondary component being a CFL SQS. 
In addition, we find that, for a $2.6\Msun$ star like GW190814's secondary component, the radius is $11.35\,{\rm km}<R_{2.6}<13.67\,{\rm km}$ and the corresponding tidal deformability is $4.6<\Lambda_{2.6}<26.3$. 
Upcoming measurements on them, when available, should shed light on the nature of compact objects with such large masses.

\section{Summary}

We have performed a Bayesian analysis on the tidal deformability observations of the gravitational-wave events GW170817 and GW190425 by assuming the merging stars are SQSs based on bag-model-like EOSs.
The results indicate that the sound speed in the SQS matter is approximately a constant close to the conformal limit of $c/\sqrt{3}$, in stark contrast with its rapid growth behavior in NS matter.
The SQSs are found to have a maximum mass of at most $M_{\rm TOV}=2.31\Msun$ at a $90\%$ confidence level; it is $2.10_{-0.12}^{+0.12}\Msun$ and $2.15_{-0.14}^{+0.16}\Msun$ for the normal and CFL matter, respectively.
Several universal relations between the observed properties of SQSs are provided.
We also find that GW190814’s secondary component could be a CFL quark star if the effective bag constant and CFL pairing are large enough.

\acknowledgments
We are thankful to Alessandro Drago, Zheng Cao, and Toru Kojo for helpful discussions. 
The work is supported by National SKA Program of China (No. 2020SKA0120300), the National Natural Science Foundation of China (grants Nos. 11873040 and 11625521), the science research grants from the China Manned Space Project (No. CMS-CSST-2021-B11) and the Youth Innovation Fund of Xiamen (No. 3502Z20206061).
This research has made use of data and software obtained from the Gravitational Wave Open Science Center (\url{https://www.gw-openscience.org}), a service of LIGO Laboratory, the LIGO Scientific Collaboration, and the Virgo Collaboration. LIGO is funded by the U.S. National Science Foundation. Virgo is funded by the French Centre National de Recherche Scientifique (CNRS), the Italian Istituto Nazionale della Fisica Nucleare (INFN) and the Dutch Nikhef, with contributions by Polish and Hungarian institutes.

\software{Bilby \citep[v0.5.5, \url{https://git.ligo.org/lscsoft/bilby/}]{2019ascl.soft01011A}, PyMultiNest \citep[v2.6, \url{https://github.com/JohannesBuchner/PyMultiNest}]{2016ascl.soft06005B}.}

\end{document}